\documentclass[12pt]{iopart}
\usepackage{bm,graphics,graphicx,epsfig,color}
\usepackage{iopams}
\begin{document}
\title{\rm \bf Testing post-Newtonian theory with 
gravitational wave observations}
\author{K.G. Arun$^1$, B.R. Iyer$^1$, M.S.S. Qusailah$^1$$^{,2}$ and\\ B.S.
Sathyaprakash$^3$}
\ead{arun@rri.res.in, bri@rri.res.in, mssq@rri.res.in, B.Sathyaprakash@astro.cf.ac.uk}
\address{$^1$Raman Research Institute, Sadashivanagar, 
Bangalore 560080, India\\
$^2$On leave of absence from Sana'a university, Yemen\\
$^3$School of Physics and Astronomy, Cardiff University, 5, 
The Parade Cardiff, CF24 3YB, UK
}
\begin{abstract}
The Laser Interferometric Space Antenna (LISA)  will observe 
supermassive black hole binary mergers with amplitude signal-to-noise 
ratio of several thousands. We investigate the extent to which such
observations afford high-precision tests of Einstein's gravity.
We show that LISA provides a unique opportunity to probe the non-linear 
structure of post-Newtonian theory both in the context of general 
relativity and its alternatives.
\end{abstract}
\maketitle
In a black hole binary, as the two holes orbit about their 
centre-of-mass, the energy and angular momentum 
from the system is dissipated into gravitational radiation. 
The radiation back-reaction force causes the two bodies to 
gradually spiral in towards each other, resulting in a strong burst 
of radiation just before they merge to form a single black hole. The 
radiation emitted at the end of the binary evolution is the primary 
target of both the ground- and space-based interferometric gravitational 
wave (GW) antennas~\cite{Thorne300Years}. It has become necessary to gain an accurate understanding 
of the late-time evolution of binaries in order to help data analysts 
in detecting the signal and measuring the parameters 
by fitting the observed signal with that expected from general 
relativity. 

In General Theory of Relativity there is no exact solution to the 
two-body problem.  In the absence of an exact solution (analytical 
or numerical) theorists have resorted to an approximate solution to 
the problem using post-Newtonian (PN) theory. Briefly, the programme 
of PN theory is the following. Let us consider a binary consisting 
of two non-spinning black holes of masses $m_1$ and $m_2$ (total 
mass $M=m_1+m_2,$ symmetric mass ratio $\eta=m_1 m_2/M^2$) 
separated by an orbital (Schwarzschild coordinate) distance $r(t).$ 
Post-Newtonian theory expresses the relevant 
physical quantities as a series in the `small' velocity parameter 
$\epsilon = 
v/c \equiv \sqrt {GM/rc^2}.$ For a system consisting of non-spinning black 
holes the only relevant quantities are the (specific) binding 
energy $E$ and the GW flux $\cal F,$ which are obtained
as perturbative expansions in $v$.
(Following the standard
convention, in units where $G=c=1$, $v^n$ corresponds to a term of order $\frac{n}{2}$-PN.)  
Currently, these expansions are
known to order $v^7 $ or 3.5PN~\cite{2PNphasing,phasing}:
\begin{equation}
E = -\frac{1}{2}\eta v^2\sum_{k=0}^{3} E_k v^{2k}, \ \ \ \ 
{\cal F} = \frac{32}{5}\eta^2 v^{10} \sum_{k=0}^{7} {\cal F}_k v^{k},
\end{equation}
where $E_k$ and ${\cal F}_k$ are the PN expansion
coefficients that are all functions of the mass ratio $\eta.$

In the {\it adiabatic approximation} one then uses the energy balance
equation, $-dE/dt = \cal F,$ to compute the evolution of the orbital 
phase $\varphi(t)$ using the following coupled ordinary 
differential equations:
\begin{equation}
\frac{d\varphi}{dt} = \omega = \frac{v^3}{M},\ \ \ \ 
\frac{dv}{dt} = \frac{dE/dt}{dE/dv} = \frac {-\cal F}{E'(v)},
\end{equation}
where $E'(v)\equiv dE/dv.$  
The phase $\Phi(t)$ of the emitted radiation at dominant order is simply
twice the orbital phase: $\Phi(t) = 2\varphi(t)$.

The phasing formula obtained by solving the above differential equations
includes different PN terms 
arising from nonlinear multipole interactions as the wave propagates
 from the source's near-zone to the far-zone~\cite{tail}.
The 1.5PN and 2.5PN term arise
 solely due to the interaction of the Arnowitt-Deser-Misner (ADM) mass of the source
and the quadrupole moment.
It is  physically  due to the scattering of 
quadrupolar waves off the Schwarzschild curvature generated
 by the source and is referred to as 
the gravitational wave  tail. 
The 3PN term includes, in 
addition to the terms at the retarded time, more interestingly 
the cubic nonlinear interactions due to the scattering of the wave tails
by the ADM mass energy of the spacetime.
The observational tests of these PN terms in effect
test the nonlinear structure of Einstein's gravity.

In the restricted PN approximation
(so called because one keeps only PN corrections to the phase of the
radiation but neglects amplitude), the response of an interferometric
antenna to the incident radiation from a source at a luminosity distance
$D_L$ is
\begin{eqnarray}
h(t) &=& \frac{4{\cal C{\cal M}}}{D_L} \left [ \pi{\cal M}F(t) \right ]^{2/3} 
\cos \Phi(t),
\end{eqnarray}
where ${\cal M}=\eta^{3/5}M$ is the so-called {\it chirp mass} of
the system, $F(t)\equiv \frac{1}{2\pi} \frac{d\Phi(t)}{dt}$ 
is the instantaneous frequency
of the radiation, $0 \le {\cal C} \le 1$ is a dimensionless 
geometric factor that depends on the relative orientation of 
the binary and the detector and its average over all orientations 
is $\overline{\cal C}= 2/5.$ The importance of including PN 
corrections in the phase, and neglecting them in the 
amplitude, was realized quite early on~\cite{Last3Min} and has led to
a lot of simplification of the data analysis problem.  For the tests 
proposed in this Letter it  may eventually  be necessary to 
incorporate these amplitude corrections~\cite{GWpol}. For the sake of simplicity, 
however, we have refrained from doing so in this work.

For our purposes it will be useful to work with the Fourier 
transform $\tilde h(f) \equiv \int_{-\infty}^{\infty} h(t)\, 
\exp(2\pi i f t) dt$ of the above signal.  Using the stationary 
phase approximation it has been shown 
that \cite{Thorne300Years,SathyaprakashAndDhurandhar}
\begin{equation}
\tilde h(f) = {\cal A}\, f^{-7/6} \exp\left [i \Psi(f) + i \frac{\pi}{4}\right ],
\end{equation}
with the Fourier amplitude ${\cal A}$ and phase $\Psi(f)$ given by
\begin{eqnarray} 
{\cal A} & = & \frac{\cal C}{D_L\pi^{2/3}} \sqrt{\frac{5}{24}} {\cal M}^{5/6},\nonumber\\
\Psi(f) & = & 2\pi f t_C + \Phi_C + \sum_k \psi_k f^{(k-5)/3}.
\label{eq:Fourier Phase}
\end{eqnarray} 
Here $t_C$ and $\Phi_C$ are the fiducial epoch of merger and the
phase of the signal at that epoch.  The coefficients in the PN 
expansion of the Fourier phase are given by~\cite{DIS3,DIS4,AISS05}:
\begin{eqnarray}
\label{eq:psikvsmass1}
\psi_k & = & \frac{3}{128\,\eta}(\pi\, M)^{(k-5)/3}\alpha_k,\\
\alpha_0 &=&1,\ \ \ \
\alpha_1 =0,\ \ \ \
\alpha_2 =\frac{3715}{756}+\frac{55}{9}\eta ,\ \ \ \
\alpha_3 =-16 \pi,\nonumber\\
\alpha_4 &=&\frac{15293365}{508032}+\frac{27145}{504} \eta+\frac{3085}{72} \eta ^2,\nonumber\\
\alpha_5 &=&\pi \left(\frac{38645}{756}-\frac{65}{9}\eta\right)
\left[1+\ln\left(6^{3/2}\pi M\,f\right)\right]\nonumber\\
\alpha_6 &=&\frac{11583231236531}{4694215680}-\frac{640}{3}\pi
^2-\frac{6848}{21}\gamma\nonumber \\
&+&\left(-\frac{15737765635}{3048192}+ \frac{2255}{12}\pi ^2\right)\eta
\nonumber\\
& + &\frac{76055}{1728}\eta^2-\frac{127825}{1296}\eta^3
-\frac{6848}{63}\ln\left(64 \pi M\, f\right),\nonumber\\
\alpha_7 &=&\pi\left( \frac{77096675}{254016}+\frac{378515}{1512}
\eta -\frac{74045}{756}\eta ^2\right).
\label{eq:psikvsmass2}
\end{eqnarray}
Unlike the other parameters $\psi_k,\,k\neq 5,6,$ which depend 
only on the masses of the system and are thus constants, 
the parameters $\psi_5$ and $\psi_6$ are {\it not} constants and 
have some $\log f$-dependence.
We treat the log-terms as constants with the justification
that the log-dependence on the frequency is weak in the relevant
 bandwidth.  Another possible choice is where the coefficients of
 the log-terms are treated as additional signal parameters. This
 choice, indeed, increases the dimensionality of the parameter space
 making the Fisher matrix highly ill-conditioned and we do not discuss
 it here.

Post-Newtonian theory has been highly successful in explaining the
decay of the orbital period in binary pulsars and in confirming the
emission of gravitational radiation by these relativistic systems
(cf.~\cite{Taylor} and references therein).
Nevertheless, the binary pulsar radio observations do not test
PN theory to a high order. This is because the typical
velocity in the most relativistic of binary pulsars is $v\sim
3 \times 10^{-3},$
which is not large enough for higher order terms to be important.

GW observation of the coalescence of binary black holes (BBH) will
provide a unique opportunity to test the PN theory to very
high orders. This is because the velocities in the system, close
to the merger, could be as high as $v \simeq 0.2$-$0.4,$ making
the highest order known PN term  $10^{12}$-$10^{14}$ times more 
important for GW observations than for radio binary 
pulsars and several tests of general relativity have already been
proposed by various authors \cite{BlanchetAndSathyaprakash,Will98,BBW}.  
 We shall show in this Letter that the brightest events 
that can be expected in the space-based Laser Interferometer Space 
Antenna (LISA) will test all the PN terms computed so far.

Our proposal to test the PN theory is the following: let us
suppose we have a GW event with a high
signal-to-noise ratio (SNR), say more than 1,000.  Once an event
is identified we suggest to fit the data to a signal wherein 
each term in the PN expansion is treated as an independent 
parameter. More precisely, instead of fitting the detector output with
a signal that depends on only the two mass parameters, 
we could fit it with
the same signal but by treating all the $\psi_k$'s as independent.
For example, if we want to test the PN theory to order 
$v^4$ then we should use a four-dimensional grid of templates consisting
of $\{\psi_0,\  \psi_2,\  \psi_3,\  \psi_4\}$ rather than the 
two-dimensional one that is used in the detection problem. This 
higher-dimensional fitting of the data with our model would measure 
each of the PN coefficients independently of the others. 
In Einstein's theory, for the case of nonspinning binaries, each of the $\psi_k$'s has a specific relationship 
to the masses, $\psi_k =  \psi_k(m_1, \, m_2),$ whereas in a different 
theory of gravity (for example, a theory in which
the graviton has non-zero mass) the relationship will be different and might involve
new parameters. Thus, the measured $\psi_k$'s 
could be interpreted, in principle, in the context of different theories of 
gravitation. 

In the case of general relativity we know that the 
$\psi_k$'s are given in terms of the masses by 
Eqs.~(\ref{eq:psikvsmass1}) and (\ref{eq:psikvsmass2}).
If general relativity (or, more precisely, the PN theory that 
approximates general relativity) correctly describes the dynamics of the 
system then the parameters must be consistent with each other within their
respective error bars. One way to check the consistency would be to
invert the relationships between the $\psi_k$'s and the masses
to obtain $m_2 = m_2^k (m_1,\, \psi_k),$ and
plot $m_2$ as a function of $m_1$ for various $\psi_k$'s,
and see if they all intersect at a common point.  If they do, then
the theory is correct to within the measurement errors, if not, the theory 
is in trouble. 
In addition to the PN theory we could also test other approximants, 
such as the P-approximant \cite{dis98} or the effective one-body 
approximation \cite{eob} that have been proposed as alternatives to 
the orbital dynamics of binary inspirals  
as also numerical relativity predictions.

Although these tests are in principle possible, an important question is
whether the various PN coefficients can really be measured
accurately enough for the test to be meaningful. We already know that 
a simpler test proposed in Ref.~\cite{BlanchetAndSathyaprakash}, in the
context of ground based detectors, requires 
events with SNRs in excess of
25. The generalized tests proposed in this Letter would require much
stronger signals, SNRs of 100 to test lower-order terms and in excess
of 1,000 to test all terms currently known. Initial ground-based
interferometers are unlikely to observe events with such large SNRs.
While some of these tests might be possible with advanced detectors,
a supermassive BBH merger in LISA is our best bet.

To test an approximation it should be possible to measure the
various PN coefficients with a good accuracy. We shall require that
the relative error in the measurement of a parameter be less than
100\%, i.e. $\Delta \psi_k/\psi_k \le 1,$  where $\Delta \psi_k$ is the
error in the estimation of the parameter $\psi_k,$ in order that its
presence is tested with confidence. 

Next, we will work out the accuracy with which we can measure the various
parameters in the standard PN theory.
Given a signal $\tilde h(f;\, \theta_k)$ that depends on the parameters $\theta_k$
the information matrix $g_{ab}$ is given by
\begin{equation}
g_{km} = \left < h_k, \, h_m \right >,
\end{equation}
where $h_m = \partial h(f;\, \theta_k) / \partial \theta_m,$ and for
any two functions $a$ and $b$ their scalar product $\left <a,\, b \right >$
is defined as:
\begin{equation}
\left <a,\, b \right > \equiv 2 \int_0^{\infty} \left [\tilde a(f) \tilde b^*(f) + 
\tilde a^*(f) \tilde b(f) \right ] \frac {df}{S_h(f)},
\end{equation}
with $\tilde a(f)$ being
the Fourier transform of $a(t).$ The quantity $S_h(f)$ is the
one-sided noise power spectral density (PSD) of the detector 
with units of Hz$^{-1};$
it is the only characteristic of the detector that enters the calculation 
of the errors in the estimation of parameters. 

The variance-covariance matrix $C_{ab}$ is nothing but the inverse 
of the information matrix, $C_{ab} = g^{-1}_{ab}.$
The diagonal components $C_{kk}$ are the variances in 
the parameters $\theta_k.$ $C_{km}/\sqrt{C_{kk}C_{mm}},$ $k\ne m,$ 
are the correlation coefficients between parameters 
$\theta_k$ and $\theta_m$, taking on values in the range $[-1,\, +1].$
\begin{figure}[b]
\vskip 6 pt
\includegraphics[width=3.0in]{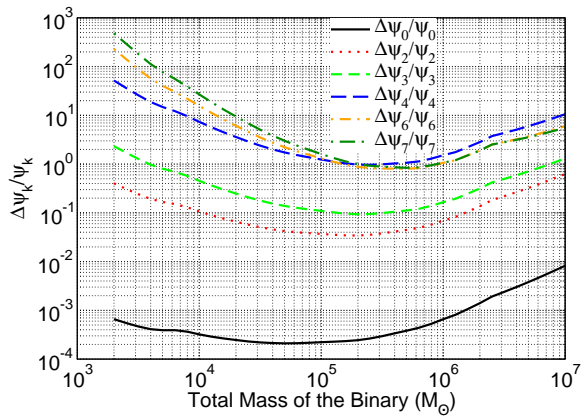}
\caption{Errors in the measurement of the PN coefficients.
The sources are optimally oriented supermassive BBH 
located at a luminosity distance of $D_L=3$ Gpc.
}
\label{fig:LISAErrors}
\end{figure}

For the binary system in question defining a new parameter vector 
${\bm \theta} \equiv \{t_C,\, \Phi_C,\, \psi_0,\,\psi_2,\,\cdots, \psi_7\}$ one finds
\begin{eqnarray}
g_{ab} & = & 4 {\cal A}^2 \int_{f_{s}}^{f_{\rm lso}} f_{ab} f^{-7/3}\frac{df}{S_h(f)},\nonumber \\
f_{ab} & \equiv & \frac{\partial \Psi(f;\, {\bm \theta})}{\partial \theta_a}
\frac{\partial \Psi(f;\, {\bm \theta})}{\partial \theta_b}.
\end{eqnarray}
Note that we have now introduced specific values for the limits in the 
integration: $f_s$ is simply the lower frequency cutoff chosen such that
the loss in the SNR due to this choice is negligible
compared to the choice $f_s=0$ and $f_{\rm lso}$ is the frequency of the 
radiation at the last stable orbit of the system which we assume
to be the value given in the test mass
approximation, namely $f_{\rm lso} = 1/(6^{3/2} \pi M).$
Using the Fourier phase given in Eq.~(\ref{eq:Fourier Phase})
it is straightforward to compute $f_{ab}:$ 
 \[\left[
\begin{array}{llllllll}
4\pi^2 x^6 & 2\pi x^3 & 2\pi x^{-2} & 2\pi   & 2\pi x & 2\pi x^2 & 2\pi x^4 & 2\pi x^5\\
2\pi x^3   & 1        & x^{-5}      & x^{-3} & x^{-2} & x^{-1}   & x        & x^2     \\
2\pi x^{-2}& x^{-5}   & x^{-10}     & x^{-8} & x^{-7} & x^{-6}   & x^{-4}   & x^{-3}  \\
2\pi       & x^{-3}   & x^{-8}      & x^{-6} & x^{-5} & x^{-4}   & x^{-2}   & x^{-1}  \\
2\pi x     & x^{-2}   & x^{-7}      & x^{-5} & x^{-4} & x^{-3}   & x^{-1}   & 1       \\
2\pi x^{2} & x^{-1}   & x^{-6}      & x^{-4} & x^{-3} & x^{-2}   & 1        & x       \\
2\pi x^{4} & x        & x^{-4}      & x^{-2} & x^{-1} & 1        & x^{2}    & x^3     \\
2\pi x^{5} & x^{2}    & x^{-3}      & x^{-1} & 1      & x        & x^3      & x^{4}
\end{array}
\right]\] 
where $x=f^{1/3}.$ 
We see that the information matrix will involve moments of the
noise spectrum of the form $\int_0^\infty f^{-j/3}\frac{df}{S_h(f)},$ where
$j$ runs from $1$ to 17. 
The elements of the information matrix, therefore, take on values in 
a very large 
range leading to a highly ill-conditioned matrix. Extreme caution should be 
exercised in computing the moments, else it is easy to end up with values 
in the covariance matrix that are negative, and even imaginary, while we 
know that the covariance matrix should be real. 
We have verified our results using three different methods for numerical
integration and found that they differed from one other by no more than 10 \%.

We assumed that LISA consists of only one interferometer 
with sensitivity as in Ref.~\cite{Cutler98,Barack&Cutler04,BBW} and  
the binary consists of two black holes of equal masses in quasi-circular
orbit and observed for one year before merger.
Fig.~\ref{fig:LISAErrors} plots the error in the various parameters 
$\psi_k$ as a function of the total mass for optimally oriented 
supermassive BBH mergers at a distance of $D_L=3$ Gpc.
We shall be interested in binaries with their total mass in 
the range $2\times10^4M_{\odot}-2\times10^7 M_\odot.$ At 3 Gpc such sources would 
produce SNR in the range 1,000-6,000.  From Fig.~\ref{fig:LISAErrors}, 
wherein the log-terms are treated as constants, it is clear that 
in the mass range $10^5-10^6$, fractional errors associated with most of
the parameters are less than 1.
Thus, LISA will provide a unique opportunity 
to test the PN and related approximations to a high degree of accuracy.

\begin{figure}[b]
\vskip 6 pt
\includegraphics[width=3.0in]{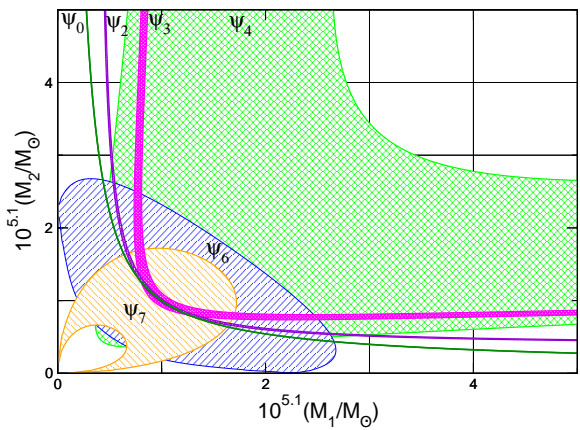}
\caption{The $m_1$-$m_2$ plane plot for different $\psi_k$'s
for the merger of a BBH of $2\times10^{5.1} M_\odot,$ at a
luminosity distance of 1Gpc.
}
\label{fig:M2vsM1}
\end{figure}

 Fig.~\ref{fig:M2vsM1} illustrates how one might test for the
 consistency of the individual masses for the specific system of
 the merger of black holes each of mass $10^{5.1}M_\odot$. We
 have chosen a system that gives the lowest errors in different
 parameters (cf.\ Fig~\ref{fig:LISAErrors}) and assumed that
 the source is at a distance of 1 Gpc.
For each parameter $\psi_k$
 we have plotted the region enclosed by the boundaries
 $\psi_k + \Delta \psi_k$ and $\psi_k-\Delta\psi_k,$ where
 $\Delta\psi_k$ is the one-sigma error in the estimation of
$\psi_k$\footnote{Since $\psi_6$ has only log-dependence on the
frequency, we
 have treated it to be a constant and have chosen $f=f_{\rm lso}$ in
 $\psi_6$  in our calculations. Other choices of the frequency would
 alter the shape of the contour corresponding to $\psi_6$ in the
 $m_1-m_2$ plane, and our choice corresponds to the one for which  $\psi_6$
will be the largest as the signal sweeps the detector band
 and leads to smallest errors.  Finally, there is no test corresponding
to $\psi_5$ in Fig~\ref{fig:LISAErrors} and
\ref{fig:M2vsM1} as a constant $\psi_5$ gets absorbed
 into a redefinition of the coalescence phase $\Phi_C$.}.
 The higher order $\psi_k$'s will have to enclose the region
 determined by, say $\psi_0$ and $\psi_2.$ This will be a stringent
 test for the various parameters and will be a powerful test
 if LISA sees a merger event with a high SNR
 of $\sim 10^4$. For binaries that merge within 1 Gpc
 the test would confirm the values of the known PN coefficients
 to within a fractional accuracy of $\sim 1.$ 

It would be interesting to ask whether the proposed tests can, in principle,
distinguish general relativity (GR) from, say a theory that
also includes a massive graviton \cite{Will98}. In this theory the 1PN
parameter $\psi_2$ is different from that in GR. Remarkably, we find that an
year's worth of  observation of BBH mergers in the mass range 
$2\times 10^{4}- 2\times 10^{7}M_{\odot}$ should be sufficient to discriminate GR from 
a massive graviton theory provided the Compton wavelength of graviton 
$\lambda_g\leq 5.5\times10^{14}-3.8\times 10^{15}$ kms.
These limits make the simplifying assumption of neglecting the as yet
uncomputed higher PN order corrections to GW phase in the massive
graviton case. (See also \cite{DE} for a discussion regarding the extent
to which GW observations can critically distinguish between different
theories of gravitation
in comparison to the binary pulsar tests.)

In the present work we have dealt with only non-spinning binaries. The spin
parameters, $\beta$ from spin-orbit coupling at 1.5PN and $\sigma$
from spin-spin coupling at 2PN, are assumed to be less significant
for these equal mass systems. 
For the unequal mass case the spin effects  are expected to be 
more important.  Orbital eccentricity, which might introduce 
systematic effects in these tests, has not been included as
we have restricted our analysis to binaries in quasi-circular orbits.
Massive graviton theories can be tested since they lead to 
a phasing formula that is structurally similar to general relativity 
but with terms modified due to the propagation delay.  Alternative theories of 
gravity, such as the Brans-Dicke theory, where the PN structure of
the phasing is different due to the presence of dipolar
radiation, may also be tested by a straightforward extension of the
above proposal.  These and other issues
will be investigated in future.
\section*{Acknowledgments}
BRI thanks the University of Wales and Cardiff, U.K. for supporting his
visit in January
2006 and  BSS thanks the Raman Research Institute, India for hospitality
during
August 2005 when part of the work was carried out.
{}

\end{document}